\documentclass[prl,twocolumn,showpacs,superscriptaddress,lengthcheck]{revtex4}

\usepackage{graphicx}
\usepackage{verbatim}
\usepackage{color}
\usepackage{amsmath}
\usepackage{microtype}
\usepackage{ulem}

\begin{document}

\newcommand{\ab}[1]{{\color{TealBlue}{[{\bf ab}: #1]}}}

\newcommand{\bra}[1]{\left\langle #1\right|}
\newcommand{\ket}[1]{\left|#1\right\rangle}
\newcommand{\braket}[2]{\left\langle #1|#2\right\rangle}
\newcommand{\com}[2]{\left[#1,#2\right]}
\newcommand{\braketop}[3]{\left\langle #1\left|#2\right|#3\right\rangle}
\newcommand{\mean}[1]{\left\langle #1 \right\rangle}
\newcommand{\trace}[2][]{{\rm Tr_{#1}}\left(#2\right)}
\newcommand{\ImaginaryPart}{{\rm Im}}
\newcommand{\RealPart}{{\rm Re}}
\newcommand{\leftexp}[2]{{\vphantom{#2}}^{#1}{#2}}
\newcommand{\leftind}[2]{{\vphantom{#2}}_{#1}{#2}}
\newcommand{\elem}{\in}
\newcommand{\rp}{\right)}
\newcommand{\lp}{\left(}
\newcommand{\lcb}{\left\{}
\newcommand{\rcb}{\right\}}
\newcommand{\rsb}{\right]}
\newcommand{\lsb}{\left[}
\newcommand{\lbv}{\left|}
\newcommand{\rbv}{\right|}
\newcommand{\lvb}{\lbv}
\newcommand{\rvb}{\rbv}
\newcommand{\bs}{\boldsymbol}
\renewcommand{\inf}{\infty}
\newcommand{\myfrac}[2]{^{#1\negthickspace\negthickspace}/_{\negthinspace#2}}
\newcommand{\mycaption}[2]{\caption[#1]{\small #1 #2}}
\newcommand{\order}[1]{{{\mathcal O}\lp#1\rp}}
\newcommand{\iohbar}{\frac{-i}{\hbar}}
\newcommand{\melem}[1]{_{#1}}
\renewcommand{\eqref}[1]{Eq.~(\ref{#1})}

\newcommand{\qrho}{\rho}
\newcommand{\crho}{\varrho}
\newcommand{\gd}{\gamma_\downarrow}
\newcommand{\gu}{\gamma_\uparrow}

\newcommand{\superop}[1]{{\mathcal #1}}
\newcommand{\sD}{{\superop{D}}}
\newcommand{\sL}{{\superop{L}}}
\newcommand{\sC}{{\superop{C}}}
\newcommand{\sT}{{\superop{T}}}
\newcommand{\sM}{{\superop{M}}}
\newcommand{\sB}{{\superop{B}}}

\newcommand{\tr}[1]{\mathbf{#1}}
\newcommand{\tU}{{\tr{U}}}
\newcommand{\tS}{{\tr{S}}}
\newcommand{\tD}{{\tr{D}}}
\newcommand{\tR}{{\tr{R}}}
\newcommand{\tP}{{\tr{P}}}
\newcommand{\tT}{{\tr{T}}}
\newcommand{\trans}[1]{^{#1}}

\newcommand{\atr}[1]{{\mathbf #1}}
\newcommand{\atU}{{\atr{U}}}
\newcommand{\atS}{{\atr{S}}}
\newcommand{\atD}{{\atr{D}}}
\newcommand{\atR}{{\atr{R}}}
\newcommand{\atP}{{\atr{P}}}
\newcommand{\atT}{{\atr{T}}}

\newcommand{\g}{g}
\newcommand{\e}{e}
\newcommand{\f}{f}

\newcommand{\wg}{\omega_\g}
\newcommand{\we}{\omega_\e}
\newcommand{\wf}{\omega_\f}
\renewcommand{\wr}{\omega_r}
\newcommand{\wm}{\omega_m}

\newcommand{\proj}[1]{\Pi_{#1}}

\newcommand{\ad}{a^\dag}
\newcommand{\ada}{a^\dag a}
\newcommand{\sz}{\sigma_0}
\newcommand{\szd}{\sigma_0^\dag}
\newcommand{\so}{\sigma_1}
\newcommand{\sod}{\sigma_1^\dag}
\newcommand{\gz}{g_0}
\newcommand{\go}{g_1}
\newcommand{\izp}{I_{0+}}
\newcommand{\izm}{I_{0-}}
\newcommand{\izpm}{I_{0\pm}}
\newcommand{\izmp}{I_{0\mp}}
\newcommand{\iop}{I_{1+}}
\newcommand{\iom}{I_{1-}}
\newcommand{\iopm}{I_{1\pm}}
\newcommand{\iomp}{I_{1\mp}}
\newcommand{\itp}{I_{2+}}

\newcommand{\epm}{\epsilon_m}
\newcommand{\Lz}{\Lambda_0}
\newcommand{\lz}{\lambda_0}
\newcommand{\Dz}{\Delta_0}
\newcommand{\Zz}{Z_0}
\newcommand{\chiz}{\chi_0}
\newcommand{\zetaz}{\zeta_0}
\newcommand{\xiz}{\xi_0}
\newcommand{\Lo}{\Lambda_1}
\newcommand{\lo}{\lambda_1}
\newcommand{\Do}{\Delta_1}
\newcommand{\Zo}{Z_1}
\newcommand{\chio}{\chi_1}
\newcommand{\zetao}{\zeta_1}
\newcommand{\xio}{\xi_1}

\newcommand{\subge}{\mathcal{E}_{\g\e}}
\newcommand{\sm}{\sigma_-}
\renewcommand{\sp}{\sigma_+}
\newcommand{\sigmaz}{\sigma_z}

\newcommand{\red}{\color[rgb]{0.8,0,0}}
\newcommand{\green}{\color[rgb]{0.0,0.6,0.0}}
\newcommand{\dkgrn}{\color[rgb]{0.0,0.4,0.0}} 
\newcommand{\blu}{\color[rgb]{0,0,0.6}}
\newcommand{\blue}{\color[rgb]{0,0,0.6}}
\newcommand{\pur}{\color[rgb]{0.8,0,0.8}}
\newcommand{\blk}{\color{black}}

\newcommand{\jcomment}[1]{\emph{\color[rgb]{0,0,0.6}(#1)}}

\title{Improved Superconducting Qubit Readout by Qubit-Induced Nonlinearities}
\date{\today}

\author{Maxime Boissonneault}
\affiliation{D\'epartement de Physique, Universit\'e de Sherbrooke, Sherbrooke, Qu\'ebec, Canada, J1K 2R1}
\author{J. M. Gambetta}
\affiliation{Institute for Quantum Computing and Department of Physics and Astronomy, University of Waterloo, Waterloo, Ontario N2L 3G1, Canada}
\author{Alexandre Blais}
\affiliation{D\'epartement de Physique, Universit\'e de Sherbrooke, Sherbrooke, Qu\'ebec, Canada, J1K 2R1}

\begin{abstract}
In dispersive readout schemes, qubit-induced nonlinearity typically limits the measurement fidelity by reducing the signal-to-noise ratio (SNR) when the measurement power is increased. Contrary to seeing the nonlinearity as a problem, here we propose to use it to our advantage in a regime where it can increase the SNR. We show analytically that such a regime exists if the qubit has a many-level structure. We also show how this physics can account for the high-fidelity avalanchelike measurement recently reported by Reed {\it et al.}  [arXiv:1004.4323v1].
\end{abstract}

\pacs{03.65.Yz, 42.50.Pq, 42.50.Lc, 74.50.+r}

\maketitle

Quantum measurements are crucial to all quantum information protocols. In solid-state systems, readout can be performed by connecting the qubits to noisy measurement electronics, such as single-electron transistors~\cite{lehnert:2003b}. Dispersive readouts based on coupling qubits to high-Q resonators are however much less disruptive as all of the energy is dissipated away from the qubit~\cite{blais:2004a}. This type of measurement leads to quantum nondemolition (QND) readout. Unfortunately, so far the typical signal-to-noise ratio (SNR) has been relatively low, requiring sometimes up to $10^6$ repetitions of the experiment to average out the noise added by the amplifiers~\cite{bianchetti:2009a,dicarlo:2009a}. 
Increasing the qubit-resonator coupling is one approach to improve the SNR as it increases the amount of information about the qubit state carried by the measurement photons. This is however at the cost of an increased Purcell decay rate~\cite{houck:2008a}, which decreases the useful integration time and in turn the SNR.
Another approach has been to exploit bifurcation in a nonlinear resonator~\cite{vijay:2009a}.
This has already led to the experimental demonstration~\cite{mallet:2009a} of single-shot readout of a transmon-type superconducting qubit~\cite{koch:2007a}. Very recently, Reed~{\it et al.} have also shown that single-shot measurement of a transmon qubit can also be realized in a \emph{linear} resonator but working at very large measurement power~\cite{reed:2010b}.

Motivated by these results, we study how the nonlinearity induced by the qubit in a linear resonator can lead to improvement of the measurement. We first explore the low driving power regime before considering the high-power regime studied in Ref.~\cite{reed:2010b}. Using a simple model, we find qualitative agreement with these experimental results. 

For simplicity, we first focus on a two-level system (TLS), of states labeled $\{\ket 0, \ket 1\}$, dispersively coupled to a microwave resonator. This system is well described by the Jaynes-Cummings Hamiltonian, expressed here in the dispersive basis ($\hbar = 1$)~\cite{boissonneault:Both}
\begin{equation}
	\label{eqn:TLSHdispersive}
	\begin{split}
		H &= \omega_r N + \frac{\omega_a-\Delta\left[ 1 - \sqrt{1+4\lambda^2(N+\Pi_{11})}\right]}{2}\sigma_z\\
		&\approx (\wr+\zeta)N  + \frac{\tilde\omega_a}{2} \sigmaz  + \chi N \sigmaz +  \zeta N^2 \sigmaz, \raisetag{16pt}
	\end{split}
\end{equation}
where $\proj{ij}=\ket{i}\bra{j}$ for the TLS, and $N=\ada$. In this expression, $\omega_r$ is the resonator frequency and $\omega_a$ ($\tilde\omega_a$) the bare (Lamb-shifted) TLS transition frequency. The qubit-resonator coupling is characterized by $\chi = g^2(1-\lambda^2)/\Delta$ the dispersive coupling strength and $\zeta = -g\lambda^3$ is a Kerr-type nonlinearity, with $g$ the bare qubit-resonator coupling strength, $\Delta = \omega_a-\omega_r$ the qubit-resonator detuning and $\lambda=-g/\Delta$. The second line of~\eqref{eqn:TLSHdispersive} is valid to fourth order in $\lambda$ and at photon numbers $\bar n\ll n_\mathrm{crit}$, where $n_\mathrm{crit}=1/4\lambda^2$ is the critical photon number~\cite{blais:2004a}. The limit $\zeta\rightarrow 0$ of the second line of~\eqref{eqn:TLSHdispersive} is the standard linear dispersive Hamiltonian~\cite{haroche:2006a}.

In this regime, because of the qubit-dependent pull of the resonator frequency $\chi\sigmaz$, a nearly resonant drive on the resonator will displace the resonator's initial vacuum state to a qubit-state dependent coherent state $\ket{\alpha_{i=0,1}}$. Homodyne measurement of the transmitted or reflected signal can then resolve these pointer states, and hence the qubit states. In Ref.~\cite{gambetta:2008a}, it was shown that the SNR of such a homodyne measurement for an integration time $T_1 = 1/\gamma_1$ is given by $\mathrm{SNR}=\eta\kappa|\alpha_1-\alpha_0|^2/\gamma_1$, where $\kappa$ is the resonator decay rate and $\eta$ is the measurement efficiency. The SNR reaches its maximal value $(\mathrm{SNR})_\mathrm{max} = 4\eta \bar n\chi/\gamma_1$, for the optimal choice $\kappa=2\chi$, where $\bar n$ is the average number of photons. In this limit, the dispersive model \eqref{eqn:TLSHdispersive} thus predicts that the SNR should increase linearly with the number of measurement photons $\bar n$.

Unfortunately, this linear increase of the SNR is not observed experimentally and this can, at least partially, be explained by the fourth order corrections in~\eqref{eqn:TLSHdispersive}. To this order of approximation, an increase of the photon number leads to a \emph{reduction} of the cavity pull $g\lambda[1-\lambda^2(\ada+1)]\sigma_z$ and hence to a loss of distinguishability of the pointer states~\cite{boissonneault:Both}. Thus, the prospects for improving the SNR in homodyne measurement of a TLS dispersively coupled to a resonator look rather unpromising.

Fortunately, most superconducting qubits are well approximated by many-level systems (MLS), often with only weak anharmonicity~\cite{valenzuela:2006a,koch:2007a,neeley:2009a,koch:2009a}, rather than by TLS. As is apparent below, it is possible in this situation for $\chi$ and $\zeta$ to have the same sign, which yields an \emph{increase} of the cavity pull with $\bar n$ and hence an improved SNR. 

As a good description of a generic superconducting qubit, we thus consider an M-level system, of states labeled $\{\ket{0},...,\ket{M-1}\}$ and with the first two states acting as logical states. The Hamiltonian of the MLS-resonator system takes the generalized Jaynes-Cummings form $H_s = H_0 + \sum_{i=0}^{M-2} g_i ( \ad \proj{i,i+1} + a \proj{i+1,i})$, where $H_0 = \wr\ada + \sum_{i=0}^{M-1} \omega_i \proj{ii}$ is the free Hamiltonian, $\omega_i$ the frequency of level $\ket i$, and $g_i$ the coupling strength between the resonator mode $a$ and the $i\leftrightarrow i+1$ MLS transition. 

In the dispersive regime, where $4 \langle a^\dag a \rangle (g_i/\Delta_i)^2\ll 1$ with $\Delta_i = (\omega_{i+1}-\omega_i) -\omega_r$,  it is convenient to approximately diagonalize $H_s$. Following Ref.~\cite{boissonneault:Both} where this was done for a TLS, we obtain to fourth order in $\lambda_i = -g_i/\Delta_i$
\begin{equation}
	\label{eqn:dispersive_hamiltonian}
	H_s^\atD \approx \tilde{H}_0 + \sum_{i=0}^{M-1} S_i \proj{i,i} a^\dag a + \sum_{i=0}^{M-1} K_i \proj{i,i} (a^\dag a)^2,
\end{equation}
where $\tilde{H}_0$ includes the Lamb shifts. In this expression, we have defined the ac-Stark ($S_i$) and Kerr ($K_i$) coefficients
\begin{subequations}
	\label{eqn:coefficient_values}
	\begin{align}
		&S_i = \lsb\chi_{i-1}(1-\lambda_i^2) - \chi_i(1-\lambda_{i-1}^2) - 2\chi_{i-1}\lambda_{i-1}^2\rsb \nonumber \\
		& \quad\quad + \tfrac14 ( 9\chi_{i-2}\lambda_{i-1}^2 - 3\chi_{i-1}\lambda_{i-2}^2 -\chi_i\lambda_{i+1}^2 + 3\chi_{i+1}\lambda_i^2 ) \nonumber \\
		& \quad\quad - g_i^{(2)}\lambda_i^{(2)} - 3g_{i-2}^{(2)}\lambda_{i-2}^{(2)}, \\
		&K_i = \tfrac14 (3\chi_{i-2}\lambda_{i-1}^2 - \chi_{i-1}\lambda_{i-2}^2 + \chi_i\lambda_{i+1}^2 - 3\chi_{i+1}\lambda_i^2) \nonumber \\
		& + (\chi_i-\chi_{i-1})(\lambda_i^2+\lambda_{i-1}^2) + g_i^{(2)}\lambda_i^{(2)} - g_{i-2}^{(2)}\lambda_{i-2}^{(2)}, \raisetag{14pt}
	\end{align}
\end{subequations}
with $\chi_i = g_i^2/\Delta_i$, $g_i^{(2)} = \lambda_i\lambda_{i+1}(\Delta_{i+1}-\Delta_{i})$, $\lambda_{i}^{(2)}=-g_i^{(2)}/(\Delta_{i+1}+\Delta_{i})$, and $\chi_i=\lambda_i=0$ for $i \notin [0,M-2]$. For $M=2$, $S_0 = -\chi_0$, $S_1 = \chi_0(1-2\lambda_0^2)$ and $K_0 = -K_1 = \chi_0\lambda_0^2$, reproducing \eqref{eqn:TLSHdispersive}.  The crucial feature of these expressions is that, contrary to a TLS, $K_0 \neq -K_1$ for a MLS. Using this state-dependent nonlinearity, we now show how to improve the readout in two ways.

\emph{Weak driving limit.} 
From the dispersive model \eqref{eqn:dispersive_hamiltonian}, it is possible to evaluate the difference in cavity pull $\delta$ for levels $\ket 0$ and $\ket 1$ which we are interested in distinguishing in a measurement. It takes the form $\delta = \chi' + \zeta'\bar n$, with $\bar n=\mean{\ada}$ and where we have defined $\chi' = S_1 - S_0$ and $\zeta' = K_1 - K_0$. We note that, although we are focusing on the first two levels, the presence of higher MLS levels is important. This is apparent in the expressions for $S_i$ and $K_i$ which involve states up to $\ket{i+2}$.  

\begin{figure}[t]
	\centering
	\includegraphics[width=0.95\hsize]{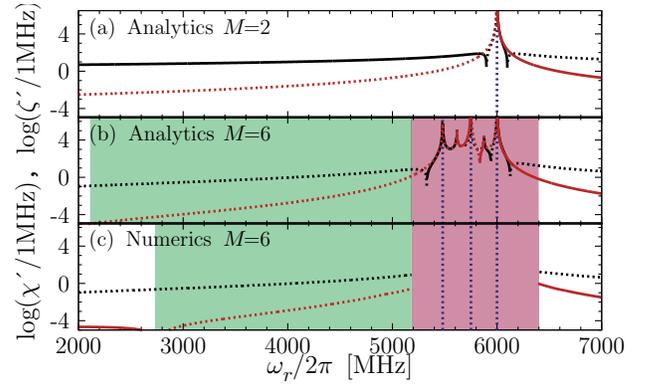}
	\caption{(color online) Analytical (a,b) and numerical (c) ac-Stark $\chi'$ (black line) and Kerr shifts $\zeta'$ [lighter (red) line] for a transmon qubit taking into account  $M=2$ (a) and $M=6$ levels  (b),(c). The parameters are chosen such that $(\omega_{10},\omega_{21},g_0)/2\pi=(6000,5750,100)$~MHz. The lines change from dotted to full when $\chi'$ or $\zeta'$ pass from negative to positive values. Vertical blue dotted lines indicate transitions $\omega_{10}$, $\omega_{21}$, and $\omega_{32}$. Lighter (green) shading indicates regions of interest, while darker (red) indicates regions where the dispersive model breaks down even at low-photon number. Values of $\chi'$ and $\zeta'$ obtained numerically are not plotted close to divergences.
	}
	\label{fig:ChiZeta}
\end{figure}
Figure~\ref{fig:ChiZeta} shows $\chi'$ and $\zeta'$ obtained from \eqref{eqn:coefficient_values} as a function of $\omega_r$ for a MLS with (a) $M=2$ and (b) $M=6$ levels. Figure 1(c) has been obtained from exact diagonalization of $H_s$ for $M=6$. For $M=2$,  sgn$(\chi')$=sgn$(\zeta')$ only in regions where the dispersive approximation breaks down while for $M>2$ this is possible in the dispersive regime (see the caption of Fig. 1). The numerical results illustrate that the analytical expressions are good approximations. We note that, while these results apply to any MLS, for concreteness we have chosen here parameters corresponding to a transmon qubit~\cite{koch:2007a}.

\begin{figure}[t]
	\centering
	\includegraphics[width=0.95\hsize]{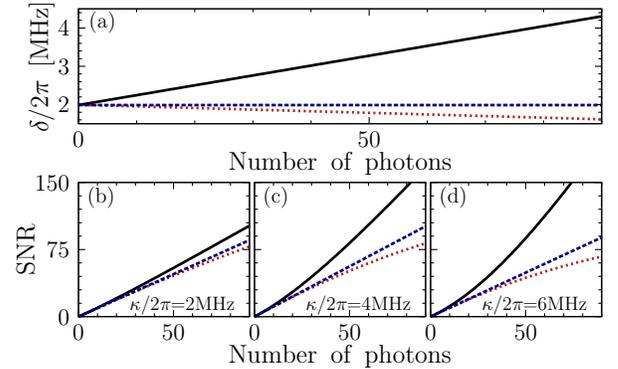}
	\caption{(color online) Cavity pull $\delta$ (a) and SNR (b)--(d), as a function of the average number of photons. The parameters are the same as in Fig~\ref{fig:ChiZeta} with $T_1 = 1~\mu$s. Dashed blue lines correspond to the second order approximation, for which the cavity pull $2\g^2/\Delta$ is constant. Full black lines (dotted red lines) are obtained for $\omega_r/2\pi = 4515(7660)$~MHz, corresponding to $n_\mathrm{crit}\sim 55(70)$ and $\mathrm{sgn}(\chi')=(-)\mathrm{sgn}(\zeta')$. These  frequencies are chosen such that $|\chi'|/2\pi = 2$~MHz.
	}
	\label{fig:snr}
\end{figure}
Figure~\ref{fig:snr}(a) shows the cavity pull $\delta$ for a cavity frequency chosen in the region where sgn$(\chi')$=sgn$(\zeta')$ (full black line) and outside of this optimal area (dotted red line).  These results are compared to the result of the second order approximation (dashed blue line). As expected, the cavity pull increases with $\bar n$ under the appropriate choice of parameters. Figures~\ref{fig:snr}(b)-\ref{fig:snr}(d) show the corresponding SNR, using the same color scheme, for $\kappa /2 \chi' = 0.5$ [\ref{fig:snr}(b)], 1 [\ref{fig:snr}(c)] and 1.5 [\ref{fig:snr}(d)], with $\kappa /2 \chi' = 1$ being the optimal choice~\cite{gambetta:2008a}.  Because of the increase of the cavity pull with $\bar n$, the full black line is always above the dotted red one.  For $\kappa /2 \chi' = 1$, an improvement of nearly $100\%$ is expected at large photon numbers. Biasing the qubit above the resonator's fundamental frequency, as is suggested here, can lead to an increase of the Purcell decay. This can however be strongly reduced by a small change of design~\cite{reed:2010a}. Finally, one could also tune the system to a point where $\chi'$ and $\zeta'$ have the same sign at the moment of measurement using a tunable resonator~\cite{palacios-laloy:2007a,sandberg:2008a}.

\emph{Strong driving limit.}
The results obtained so far relied on \eqref{eqn:dispersive_hamiltonian}, which is valid only below $n_\mathrm{crit}$. However, the nontrivial state dependence of the nonlinearity $K_i$ should extend well beyond the dispersive regime. To explore this, we take advantage of the block diagonal structure of the Hamiltonian $H_s$. There each block of $H_s$ corresponds to a fixed number $n$ of qubit-resonator excitations and is spanned by $\mathcal{E}_n = \{\ket{n,0},...,\ket{n-M+1,M-1}\}$. With the relevant $M$ being at most $\sim 10$ in practice, we can diagonalize each block numerically (or analytically for $M\le4$) for arbitrary $n$. In this way, we obtain the dressed energies  $\bar{E}_{n,i}$ and states $\ket{\overline{n,i}}$, where $\bar{E}_{n,i}$ is the energy of the eigenstate closest to a Fock state with $n$ photons and MLS state $\ket i$. From these expressions, we find the effective resonator frequency $\omega_{ri}(n) = \bar{E}_{n+1,i} - \bar{E}_{n,i}$. This frequency depends in a nonlinear way on the MLS state-dependent average photon number $n_i$. In steady state and in the absence of qubit transitions, $n_i$ is given by the measurement drive amplitude $\epsilon$ and frequency detuning relative to the effective resonator frequency
\begin{equation}\label{eq:ni}
	n_i(\epsilon,\omega_m) = \frac{\epsilon^2}{[\omega_{ri}(n_i) - \omega_m]^2 + [\kappa/2]^2},
\end{equation}
with $\omega_m$ the measurement frequency. We solve \eqref{eq:ni} iteratively to find $n_i$ and $\omega_{ri}$ as a function of $\epsilon$ and $\omega_m$. 

\begin{figure}[t]
	\centering
	\includegraphics[width=0.95\hsize]{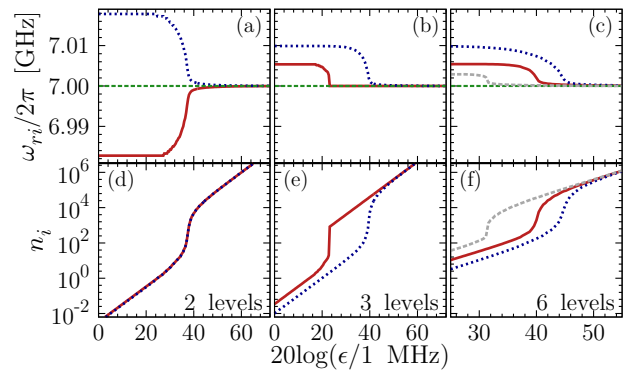}
	\caption{Effective resonator frequency $\omega_{ri}$ (a)--(c) and mean photon number $n_i$ (d)--(f) for $i=0$ (full red lines), $i=1$ (dotted blue lines), and $i=2$ [dashed gray lines, (c),(f)] as a function of the measurement power. Panels (a,d), (b,e) and (c,f) are for $M=2$, 3, and 6 respectively. In (a)--(c), the dashed green horizontal line is $\omega_r$. The parameters are the same as in Fig~\ref{fig:ChiZeta}. For clarity of presentation the panels have different horizontal scales.}
	\label{fig:trace}
\end{figure}
\begin{figure}
	\centering
	\includegraphics[width=0.95\hsize]{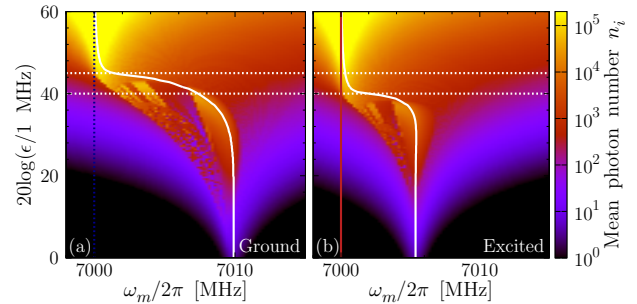}
	\caption{Mean photon number $n_i$ for $i=0$ (a) and $i=1$ (b) as a function of the measurement frequency and power. The full white lines are $\omega_{ri}(n_i)$ for $\omega_m/2\pi=\omega_r/2\pi=7$~GHz. The solid red and dotted blue vertical lines indicate the measurement frequency used in Fig~\ref{fig:trace}. The parameters are the same as in Fig~\ref{fig:ChiZeta}. The horizontal dotted lines delimitate the regime of measurement power where $n_1-n_0$ is maximum.}
	\label{fig:densityplot}
\end{figure}

In Fig.~\ref{fig:trace}, both $\omega_{ri}$ [\ref{fig:trace}(a)--\ref{fig:trace}(c)] and $n_i$ [\ref{fig:trace}(d)--\ref{fig:trace}(f)] are plotted as a function of measurement power for $M=2,3,6$ respectively. This is done for $\omega_m = \omega_r$ as in  Ref.~\cite{reed:2010b}. For all values of $M$, the effective resonator frequency approaches its bare value $\omega_r$ at large power. This is expected because, at this point, $\langle N \rangle \gg \langle \sqrt{N} \rangle$ in \eqref{eqn:TLSHdispersive} and the cavity responds classically~\cite{fink:2010a}.  Since $K_0=-K_1$ for $M=2$, this classical crossover occurs at the same input power for both qubit states. We note that this crossover happens in an avalanche manner, with each additional photon bringing $\omega_{ri}(n_i)$ closer to $\omega_r$ and facilitating the addition of more photons. As can be seen in Fig.~\ref{fig:trace} (b,c), for $M=3$ this avalanche occurs at a state-dependent power, the behavior changing only quantitatively for $M>3$. With $\omega_{m}=\omega_r$, we thus expect an abrupt change in the average photon number in the resonator at a power that is MLS state dependent. This is illustrated in Fig.~\ref{fig:trace}(f) where for $M=6$ there is a range of $\sim$ 5 dB in measurement power where $n_0$ and $n_1$ differ significantly and by as much as $\sim 10^5$ at the optimal driving power. This large separation of the S curves, much larger than typical amplifier noise, leads to single-shot readout of the qubit~\cite{reed:2010b}. As shown by the gray dashed line in Fig~\ref{fig:trace}(f), pumping the $1\leftrightarrow 2$ transition before readout~\cite{mallet:2009a,reed:2010b} could also help the measurement work at lower power. Finally, Fig.~\ref{fig:densityplot} shows the full power versus measurement frequency dependence of $n_i(\epsilon,\omega_m)$ obtained from \eqref{eq:ni}. In this plot, the full white lines correspond to $\omega_{ri}$ shown in Fig.~\ref{fig:trace}(c). Although Fig.~\ref{fig:densityplot} shows qualitative agreement with the results of Ref.~\cite{reed:2010b}, a quantitative comparison would require proper modeling of the 4-qubit device used in Ref.~\cite{reed:2010b}.

\begin{figure}[t]
	\centering
	\includegraphics[width=0.95\hsize]{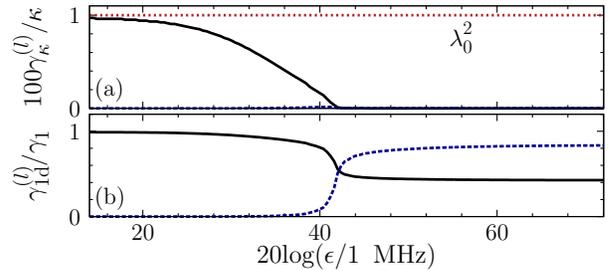}
	\caption{Qubit relaxation (full black lines) and leakage rate (dashed blue lines) due to Purcell effect~(a) and bare qubit decay~(b) as a function of the measurement power. The dotted red line in~(a) is the expected value for $\gamma_\kappa$ at low power. The parameters are the same as in Fig~\ref{fig:ChiZeta}.}
	\label{fig:gamma_eff}
\end{figure}

While the dispersive measurement at low-photon number of Fig.~\ref{fig:snr} is expected to be QND, this is not the case for the single-shot high-power measurement. To evaluate the QND character of this avalanche readout, we estimate how the presence of the measurement photons changes the relaxation and excitation rates of the qubit, as well as causes leakage outside of the logical subspace $\{\ket{0},\ket{1}\}$. Figure~\ref{fig:gamma_eff}(a) shows the Purcell decay rate $\gamma_\kappa/\kappa \approx \lvb\bra{\overline{n,0}}a\ket{\overline{n,1}}\rvb^2$ and corresponding leakage rate $\gamma_\kappa^l/\kappa \approx \sum_{i\ne0,1} \lvb\bra{\overline{n-i,i}}a\ket{\overline{n,1}}\rvb^2$ as a function of measurement power. At low power, we find the expected result $\gamma_\kappa/\kappa = \lambda_0^2$~\cite{houck:2008a}. For large photon number, $n+1 \approx n$ and losing a photon through resonator decay does not significantly change the qubit states. The Purcell decay rate thus goes down with measurement power and does not affect the QND character. To evaluate how the qubit dressing changes pure relaxation, Fig~\ref{fig:gamma_eff}(b) shows the rates for dressed decay $\gamma_{1d}/\gamma_{1} \approx \lvb\bra{\overline{n,0}}\Sigma_-\ket{\overline{n,1}}\rvb^2$ and leakage $\gamma_{1d}^{l}/\gamma_{1} \approx  \sum_{i\ne0,1}\lvb\bra{\overline{n-i,i}}\Sigma_-\ket{\overline{n,1}}\rvb^2$, where $\Sigma_- = \sum_{i=0}^{M-2} \tfrac{g_i}{g_0}\proj{i,i+1}$. The participation of the higher transmon states reduces the decay rate $\gamma_{1d}$ from $\ket{1}$ to $\ket{0}$ as measurement power is increased. However, decay of the bare higher states increases the leakage rate, and the total error rate $\gamma_{1d}+\gamma_{1d}^l$ is larger than $\gamma_1$.

Finally, dressed-dephasing $\gamma_\mathrm{d}$ due to noise responsible for dephasing of the bare qubit states can also cause transitions between the dressed-states~\cite{boissonneault:Both,wilson:2010a}. For concreteness, we consider dephasing due to charge noise on a transmon, but the model can be adapted to any source of dephasing. Following Ref.~\cite{boissonneault:Both}, this contribution can be evaluated as
$\gamma_\mathrm{d}/\gamma_\varphi \approx \lvb\bra{\overline{n+1,0}} \Sigma_z\ket{\overline{n,1}}\rvb^2 S(\overline{\Delta}_{10})/S(1\mathrm{~Hz})$, where $\Sigma_z = \sum_{i=0}^{M-1} \proj{i,i} \epsilon_i/\epsilon_1$, with $\epsilon_i$ the charge dispersion of level $i$~\cite{koch:2007a} and $S(\overline{\Delta}_{ij})$ the spectrum of charge noise evaluated at the dressed qubit-resonator detuning. In the same way, the leakage rate is $\gamma_\mathrm{d}^l/\gamma_\varphi \approx \sum_{i\ne 0,1}  \lvb\bra{\overline{n+1-i,i}} \Sigma_z\ket{\overline{n,1}}\rvb^2 S(\overline{\Delta}_{1i})/S(1\mathrm{~Hz})$. 

We note that, even assuming $1/f$ charge noise which would be $10^{9}$ times smaller at 1~GHz than at 1~Hz, dressed dephasing can be important for these large photon numbers. Indeed, for the transmon, the charge dispersion $\epsilon_i$ --- and therefore the susceptibility to charge noise --- increases exponentially with $i$, reaching $\epsilon_6/\epsilon_1 \sim 10^{6}$ for 6 levels. Although a quantitative analysis requires a better understanding of the noise spectrum at microwave frequencies, our numerical analysis with $1/f$ noise suggests $\gamma_d/\gamma_\varphi$ ranging from $1$ to $10^3$ and $\gamma_d^l/\gamma_\varphi$ from $10$ to $10^4$ depending on the parameters and the number of levels.

The loss of the QND aspect in such a high-fidelity readout is not expected to be an issue in the measurement of the final state of a quantum algorithm. However, reduction in the QND character is problematic for tasks such as measurement-based state preparation, quantum feedback control and quantum error correction.

In summary, we have shown that for a qubit with $M>2$ levels dispersively coupled to a resonator, the qubit-induced nonlinearity of the resonator depends in a nontrivial way on the qubit state. This can be exploited to increase the SNR ratio in a QND measurement at low-photon number and captures the essential aspects of the high-fidelity non-QND measurement recently reported~\cite{reed:2010b}.

\begin{acknowledgments}
We thank the Yale circuit QED team for discussion of the results of Ref.~\cite{reed:2010b} prior to publication. M. B. was supported by NSERC;  J. M. G. by CIFAR, MITACS, MRI and NSERC; A. B. by NSERC, the Alfred P. Sloan Foundation and CIFAR.
\end{acknowledgments}
\emph{Note added.} -- Theoretical modeling of the high-fidelity readout has also been reported by Bishop \emph{et al.} (following Letter)~\cite{bishop:2010a}.

\end{document}